\documentclass[12pt]{article}

\usepackage{latexsym}

\date{}

\begin{document}
\title{{\hfill{\normalsize ILAS 1/FM/2000}}\\ 
Two-dimensional conformal field theory and beyond. 
Lessons from a continuing fashion\footnote{Subject of a
Colloquium talk at the International School for Advanced Studies
(SISSA), Trieste, December $1999$ and of an Invited Lecture at the
$4^{th}$ General Conference of the Balkan Physical Union, Veliko
Turnovo , Bulgaria, $22$-$25$ August, $2000$}}

\author{I. T. Todorov \thanks{On leave of absence from the 
Institute for Nuclear Research and Nuclear Energy,
Tsarigradsko Chaussee $72$, BG-$1784$ Sofia, Bulgaria; e-mail:
todorov@inrne.bas.bg}\,\\
{\footnotesize 
Laboratorio Interdisciplinare per le Scienze Naturali e
Umanistiche,}\\ 
{\footnotesize 
SISSA-ISAS, via Beirut 2-4, I-34014 Trieste, Italy}
}
\maketitle

\begin{abstract}
Two-dimensional conformal field theory (CFT) has several sources:
the search for simple examples of quantum field theory, the
description of surface critical phenomena, the study of
(super)string vacua.

In the present overview of the subject we emphasize the role of
CFT in bridging the gap between mathematics and quantum field
theory and discuss some new physical concepts that emerged in the
study of CFT models: anomalous dimensions, rational CFT, braid
group statistics.

In an aside, at the end of the paper, we share the misgivings,
recently expressed by Penrose, about some dominant trends in
fundamental theoretical physics.
\end{abstract}


\sloppy
\renewcommand{\baselinestretch}{1.3} %
\newcommand{\sla}[1]{{\hspace{1pt}/\!\!\!\hspace{-.5pt}#1\,\,\,}\!\!}
\newcommand{\db}{\,\,{\bar {}\!\!d}\!\,\hspace{0.5pt}}
\newcommand{\partb}{\,\,{\bar {}\!\!\!\partial}\!\,\hspace{0.5pt}}
\newcommand{\dsla}{\partb}
\newcommand{\eql}{e _{q \leftarrow x}}
\newcommand{\eqr}{e _{q \rightarrow x}}
\newcommand{\ite}{\int^{t}_{t_1}}
\newcommand{\itz}{\int^{t_2}_{t_1}}
\newcommand{\itd}{\int^{t_2}_{t}}
\newcommand{\lfrac}[2]{{#1}/{#2}}
\newcommand{\dV}{d^4V\!\!ol}
\newcommand{\ben}{\begin{eqnarray}}
\newcommand{\een}{\end{eqnarray}}
\newcommand{\la}{\label}

\bigskip
\section{Introduction. Reconciling mathematics and physics}

Two-dimensional conformal field theory ($2D$ CFT) has been with us
for over 30 years. It became a main stream topic since the
breakthrough  1984 paper of Belavin, Polyakov and Zamolodchikov
\cite{BPZ}. The present overview, addressed to a broad audience,
deals with some general aspects of this development. We offer a
(necessarily subjective) answer to the question: what should a
theoretical physicist or mathematician interested in quantum field
theory (QFT) retain from it?

It seems appropriate to begin with the role $2D$ CFT is playing in
bridging the gap between mathematics and QFT\footnote{A source of
inspiration for me in this respect has been the essay of Werner
Nahm \cite{Nahm}. It belongs to one of a number of collections of
reflections on physics and mathematics at the turn of the century
(see also \cite{G}-\cite{KZ}).}. First, some general remarks about
the affair between mathematics and physics.

Archimedes, one of the greatest mathematicians of antiquity, also
was the greatest physicist of his time. Newton, the founder of
modern physics, created on the way the calculus. After him,
classical mechanics was developed by mathematicians: Euler,
Lagrange, Poisson, Hamilton, Jacobi.... The separation between
mathematics and (theoretical) physics only took place during the
$19^{th}$ century, accompanying the creation of classical
electrodynamics. Still, in the first quarter of the $20^{th}$
century mathematicians and physicists could compete on a common
ground. The three revolutions of the "golden age in physics" were
made easier by preceding developments in mathematics: Poincar\'e,
following the work of Lorentz, anticipated the creation of special
relativity (see, e.g., \cite{P}); Einstein's theory of gravity was
based on Riemannian geometry (and the action giving rise to the
correct general relativistic equations of motion was written down
by Hilbert \cite{T1} ); if quantum mechanics -- in the form which
Dirac gave to it -- was gradually understood as a deformation of
classical mechanics -- see \cite{Kont} and references therein, this
was made possible by the preceding mathematical development of
the Hamiltonian formalism. Stone-von Neumann theory of operators in
a Hilbert space was stimulated by -- and serves as language of --
quantum mechanics.

In the words of Nahm \cite{Nahm} "a theoretical physicist who
looks back to the beginning of the past century has reasons to be
humble". Indeed, the remaining three quarters of the $20^{th}$
century were devoted to a not fully successful effort to combine
two of these revolutions, special relativity and quantum
mechanics, into  quantum field theory. (Well in line with Michael
Atiyah's observation that when a theoretical physicist can not
solve a problem he goes for the next more difficult one, in the
past 20 years many of us are daydreaming to put together quantum
mechanics and general relativity in a "quantum gravity" within
(super)string theory.) Perhaps, the relatively slow progress in
quantum field theory (as compared to quantum mechanics ) can be
accounted for by the fact that classical field theory, the
dynamics of a system with infinitely many degrees of freedom, has
not undergone the same type of mathematical elaboration (and
has not reached the same level of maturity) as classical
mechanics.

In fact, it is probably not an accident that the parting of ways
between mathematics and physics roughly coincided with the first
attempts to create quantum electrodynamics in the late twenties.
Freeman Dyson was hardly exaggerating when, in his 1972 address to
the American  Mathematical Society, he deplored the divorce
between mathematics and physics over the issue of quantum field
theory.

The reasons  for such an alienation were multifold. Even during
the years of relative harmony, in the beginning of the century,
physicists did not appear to appreciate "pure mathematics".
Mittag-Leffler failed, in spite of his efforts, to get a Nobel
prize for Poincar\'e (see \cite{N}). Later, in the words of Res
Jost, "...under the demoralizing influence of quantum field
theoretic perturbation theory [infested with divergences], the
mathematics required for a theoretical physicist was reduced to a
rudimentary knowledge of the Latin and Greek alphabet." The
attempt of axiomatic field theory to establish some order in the
domain were met with contempt by "main stream" theorists.
Mathematics, on the other hand, had an inward development
personified by Bourbaki. Two leaders of the group, Andr\'e Weil
and Jean Dieudonn{\'e}, ventured to affirm that XX century
mathematics would not suffer the influence of physics
\footnote{Cited by Pierre Cartier in "Andr\'e Weil (1906-1998):
adieu \`a un ami", S\'eminaire de Philosophie et de
Math\'ematiques, \'Ecole Normale Sup\'erieure, 1998 }.

Signs of a changing attitude, on both sides of the fence,
started to be visible in the 1950's. A new brand of mathematical
physicists (Bogolubov, Faddeev, Haag, Kastler, Wightman -- to name
a few) tried to translate the problems of QFT into a proper
mathematical language. The task was not easy. In fundamental
physics to ask the right question you need to anticipate the
answer. The problems of QFT are very hard indeed. When a
mathematician is confronted with an un-manageable problem he looks
for a simpler one of a similar nature that may help him gain some
insight. Among physicists such an attitude is often viewed with
disrespect. It took some 12 years before the first interesting
example of a $2D$ QFT (Thirring 1958 \cite{Thir1}; see also
\cite{Thir2}) originally proposed as a toy model for the then
popular Heisenberg's $4$-fermion "Urmaterie" theory, was truly
appreciated and put to use by theorists. Physicists in the
1960's were working hard on $4$-dimensional current algebras, a
subject which has yet to become a mathematical theory, neglecting
to look first at the much simpler $2$-dimensional case. It was not
until 1972 (see \cite{AFZ}) that the current algebra inherent to
the Thirring model was made explicit (5 years after Kac and Moody
discovered their non-abelian current algebras). In the 60's
\cite{Thir1} was not cited in the quite extensive bibliography of
Schweber's "Introduction to Relativistic Quantum Field Theory",
1964 -- not even in Thirring's own book with Henley ("Elementary
Quantum Field Theory", 1962); a notable exception: former
Schwinger's student K. Johnson devoted a paper to the model in
1961 (it was he who taught it later to K. Wilson).

For a time there were independent parallel developments in
mathematics and physics (connections on fiber bundles - Yang-Mills
gauge fields, Kac-Moody algebras - current algebras, Atiyah-Singer
index theory - Adler-Bell-Jackiw anomalies).

Meanwhile, physicists got a better excuse to study $2D$ CFT: it was
realized \cite{K} that the critical Ising model is described by a
simple conformally invariant euclidean QFT (it has been gradually
understood -- starting with a 1953 paper by Abdus Salam on
super-conductivity -- that euclidean QFT can serve the description
of statistical mechanics and condensed matter systems). By that
time string theory has made its appearance (yet as a theory of
strong interaction) and it was recognized 
(by A. Galli in 1970) that $2D$ CFT is an essential ingredient. 
The ground was prepared for a renewed interaction between mathematicians
and physicists. Near the end of the century it resulted in an express
effort by mathematicians to learn QFT \cite{D}.

\section{What are we learning from $2D$ CFT?}

\subsection{Anomalous dimensions}

Equal time canonical (anti)commutation relations, $[\varphi(t,{\bf
x}),\varphi(t,{\bf y})]=i\delta({\bf x}-{\bf y})$ for a neutral
scalar field , $[\psi^{*}(t,{\bf x}), \psi(t,{\bf
y})]_{+}= \delta({\bf x}-{\bf y})$ for a Weyl spinor field,
determine the corresponding dimensions as ${D-2\over 2}$ for a scalar
field, ${D-1\over 2}$ for a spin ${1\over 2}$ field. The very
formulation of these relations requires the existence of sharp
time smeared fields, $\psi(t,f)= \int \psi(t,\xi)f(\xi)d\xi$,
$[\psi(t,f)^{*},\psi(t,f)]_{+}= \int\! \mid\! f(\xi)\!\mid^{2}
d\xi$; here f is a (smooth, falling at infinity) test function.

This property is questioned by axiomatic field theory. Haag's work
of 1955 led to the understanding that interacting fields can not
be viewed as a sharp time operator valued distributions (in the
sense of Schwartz): one also needs smearing in the time variable. This is a
manifestation of the subtleties
inherent in infinite dimensional systems. In quantum mechanics of
a system with a finite number of degrees of freedom the canonical
commutation relations $[q_{j},p_{k}]= i\delta_{jk}$ are maintained
in an interacting theory. Physicists tend not to believe in
mathematical subtleties unless they are forced to, so they ignored
for a time Haag's theorem. (Gelfand used to say that a physicist's
behaviour towards mathematics is similar to the attitude of an
intelligent thief towards the criminal law: he studies it just as
necessary to avoid punishment.) The punishment came from $2D$ CFT
models. Anomalous dimensions had to be discovered twice before
being taken seriously by the QFT community: first, they appeared
in the critical Ising model of a $2D$ spin system (Onsager, 1944,
interpreted in QFT terms in \cite{K}) in which the magnetization
(scalar) field $\sigma(x)$ has dimension ${1\over 8}$; secondly,
they were present from the outset in the Thirring model
\cite{Thir1} but only entered mainstream QFT when identified by
Wilson \cite{W} in his study of operator product expansions. 
There, the anomalous dimension of the
basic field $\psi$ (of canonical dimension (${1\over 2}$,0)) is
proportional to the square of the coupling constant. For those who
suspected that it was a peculiar $2D$ effect, recent work on $N=4$
(extended) super-symmetric Yang-Mills theory
\cite{Eden, Bianchi} demonstrates its relevance in
conformally invariant $4D$ QFT as well.

\subsection{RCFT (Rational Conformal Field Theories)}

Schroer, 1974, and L\"uscher and Mack, 1976, realized that
Wightman axioms plus scale invariance and the existence of a
(traceless) stress-energy tensor $T$ imply the Virasoro algebra
$Vir$ for the modes  $L_{n}$ of the chiral component of $T$:

\ben T(z)= \sum L_{n}z^{-n-2}\,, \qquad
\left[L_{n},L_{m}\right]=(n-m)L_{n+m} + {n(n^2 - 1)\over 12}\, c\,
\delta_{n,-m} \,\, .\een

Mack and L\"uscher, in their famous unpublished paper of 1976,
further demonstrated that the smallest positive value of the
central charge $c\,$ for which the representation of $Vir$ is
unitary is $\, c= {1\over 2}\,$ and that there is a gap between
$c={1\over 2}$ and the next (bigger) unitary value of $c$. This
was a rare moment when further progress in physics had to wait for
a substantial development in mathematics and such a development
did come 3 years later with the computation of Kac determinant
\cite{Kac}. It took another 5 years (and the help of two
mathematicians, Feigin and Fuchs) for physicists to digest this
result before the appearance of, perhaps, the most important paper
in the field, \cite{BPZ}. The authors found that for each of the
series of values of the central charge $c=c_{p,p^\prime}= 1 -
6{(p-p^\prime)^2\over pp^{\prime}}$, $p$, $p^\prime$ -- coprime
positive integers, there is a finite set
$\left(\Delta_{r,s}(p,p^\prime)\right)$ of rational conformal
dimensions (integer multiples of ${1\over 4pp^\prime}$) for which
the Kac determinant  vanishes, such that the corresponding primary
conformal fields $\phi_{rs}$ span an operator product algebra. In
other words, for each pair $(p,p^\prime)$ of different positive
integers one defines an RCFT -- i.e., a $2D$ CFT with a finite number
of super-selection sectors with ground states labeled by
$c_{p,p^\prime}$ and $\Delta_{r,s}(p,p^\prime)$. The next step was
to demonstrate \cite{FQS} that the subset of RCFT with $p^\prime =
p+1\,,\ \Delta_{r,s}= {\left[(p+ 1)r - ps \right]^2 -
1 \over 4p(p + 1) }$, $1\le r \le p-1$, $1\le s\le p$, $p= 3,4,...
$ ($r-s\,$ even ) exhaust the unitary minimal models (and in fact, all
unitary irreducible representations of $Vir$ with $c<1$). It gave
rise to a general study of RCFT which, for $c>1$, is still not
complete. (For a review of the subject including a proof of
L\"uscher-Mack theorem -- see \cite{FST}. Gawedzki's lectures in
\cite{D} provide a recent survey oriented towards mathematicians.)

There are several reasons making RCFT interesting. \\ (i) It
provides a non-trivial family of theories with a finite number of
super-selection sectors. These include not only QFT with finite
gauge group  (of the first kind) -- cf. \cite{BMT}, but also
theories which do not fit in the Doplicher-Roberts framework
\cite{DR} and whose understanding requires a generalization of the
concept of group (e.g., a quantum matrix algebra 
at a root of unity -- see \cite{FHIOPT} and references therein). \\ 
(ii) They give rise to rational conformal dimensions
and quantized charges which may serve to describe observed
excitation of fractional quantum Hall plateaux (see, e.g.,
Fr\"ohlich and Pedrini in \cite{KZ} and references therein). \\
(iii) Vacuum expectation values of local observable fields are
rational functions of the space-time coordinates, a property which
may be also relevant for $4$-dimensional CFT -- see \cite{NT}.

One may speculate that RCFT suggests a pattern that could explain
observed charge quantization in a realistic QFT.

\subsection{Braid group statistics}

It took a surprisingly long time after the first quantum -- Bose and
Fermi -- statistics were discovered before the possibility for more
general statistics was realized.

The symmetry of an $1$-component wave function
$\psi(x_{1},...,x_{n})$ is described by either of the
$1$-dimensional representations of the group $S_{n}$ of permutations
giving  rise to Bose and Fermi statistcs. Particles with internal
quantum numbers  have multi-component wave functions that may
transform under higher dimensional irreducible representations
(IR) of $S_{n}$. It was proven by Doplicher and Roberts \cite{DR}
in 1990, within Haag's algebraic approach to local quantum theory
\cite{H}, that for $D= 4$ (or higher) space-time dimensions these
are the most general possibilities  (at least in the absence of
massless particles); moreover, the symmetry and the composition
law of super-selection sectors is governed by a compact gauge
group of the first kind  and the tensor product decomposition of
its IR. The simple observation  that in a $2D$ space, say in the
complex plane, the manifold of non-coinciding points,
$Y_{n}=\left((z_{1},...,z_{n} \in C^{n}; i\ne j \Longrightarrow
z_{i}\ne z_{j}) \right)$ is not simply connected suggests that the
wave function $\psi(z_{1},...,z_{n})$ may be multivalued and
correspond to a braid group statistics. This observation 
has made its way in the physics literature in several instalments
\cite{New, FRu, LM}. Braid group statistics are indeed realized in $2D$ CFT
(the place of wave functions being taken by CFT correlation
functions) -- the reader will find a systematic exposition and
further references in \cite{FG}; for a brief overview -- see Sec.1
of \cite{TH}. This is not just a theoretical curiosity: anyon
statistics -- corresponding to one dimensional representations of
the braid group  -- is realized by quasi-particles in the
fractional quantum Hall effect (for a recent survey and references
to original work -- see, e.g., Fr\"ohlich and Pedrini in
\cite{KZ}). Non-abelian braid group statistics have been also
proposed (see \cite{CGT, FPSW} and references therein) for
some exotic quantum Hall fluids (corresponding to second Landau
level states).

To sum up: even ignoring (as we did so far) the most popular
appearance of $2D$ CFT -- its more speculative application to string
theory -- we see that it is important as \\ -- a field of common
interest, a cross-point, for physicists and mathematicians;
\\ -- a source and a test-ground for new concepts in QFT; \\
-- the main tool for describing the universality classes of $2D$
critical phenomena.

The present overview being rather one-sided (even within the
stated restrictions) a newly interested reader is addressed to 
\cite{FMS} that appears as a standard text-book  on $2D$ CFT.

\section{A new set of values in mathematical physics?}

In a sense, the gap between mathematics and physics has been
bridged too well: a new brand of mathematical physics has emerged
with little or no empirical ground. Happily, this had led to
important new insights in problems of pure mathematics (vertex
algebras \cite{BOR}; knot invariants, low dimensional topology,
mirror manifolds -- for a sample of early accounts of these
developments, see \cite{Jones, Yau, Witten}). It is a
common place to say that mathematics is applied to physics. A flow
of applications of physical ideas to inciting a breakthrough in
long standing mathematical problems is a novel phenomenon. It
provides a raison d'\^etre to the bold attempt to achieve
an ultimate synthesis of quantum theory and gravity which has so
far no phenomenological implications. But is it enough to justify
a monopoly of string theory at the front line of theoretical
physics? At this point I am inclined to share the reservations
expressed by Roger Penrose \cite{Penrose} who suggests that the
internet/e-mail revolution in communications helps impose
everywhere the dominant trends of fundamental physics even in
cases (he cites supersymmetry  and inflation theory) in which
these trends have no observational basis. In fact, a closer look
may cast doubt on some of the pretenses of string theory as far as
its applications to pure mathematics go. For instance, the first
step in discovering new knot invariants was made by V. Jones (see
the articles of Birman and Jones in \cite{Jones}) as an outgrowth
of his study of subfactors (a factor is a von Neumann algebra with
a trivial center). Or, this work was stimulated by the
Haag-Kastler "algebraic" approach to local quantum physics (see
\cite{H}), an approach that never enjoyed the popularity of string
theory.

String theory used to be advertised as a theory free of
divergences. Divergences, however, reappeared in it together with
the appearance of higher dimensional objects ("branes"). On the
other hand, the renormalization program of good old quantum field
theory acquired respectability in the hands of Connes and Kreimer
(see \cite{Connes} and references to earlier work cited
there) who related it to a Hopf algebra structure inherent to
non-commutative geometry and to the Riemann-Hilbert problem.

The $D$-branes serve as a tool to establish relations among the five
consistent superstring theories in $10$ dimensions suggesting the
existence of a single parent theory. No clue is given, however, how to
select the ground state, the "string vacuum". This leaves us with a vast
variety of models thus rendering the uniqueness claim rather academic.

The strive for an all embracing fundamental theory is as old as
natural philosophy and has been motivating the greatest physicists
of all times. But, if the skepticism accompanying Einstein's
solitary efforts to create a "unitary theory" may have been
somewhat shortsighted, the show of euphoria at each new turn of
string theory is hardly more justified. Now, string theory has
borrowed some of the ideas of Einstein's time: the Kaluza-Klein
compactification, the role of a skewsymmetric "$B$-field" in the
geometry of space-time (in his last attempt to construct a unified
field theory Einstein has used a metric $g_{\mu\nu}$ with an
antisymmetric part). It does have a healthy mutually beneficial
contact with modern mathematics. (As we have seen, there
are other, much less fashionable branches of QFT, which share this
quality.) What is not healthy is to claim a monopoly on fundamental
theoretical developments. It is becoming increasingly difficult
for a student of QFT  to find a decent job unless he is following
the latest fad in string theory. No doubt, it would be even more
stupid to try to forbid superstrings. Will scientists and science
managers be wise enough to avoid this everything-or-nothing
attitude, and not put all their eggs in a single basket?

\section*{Acknowledgments}

I would like to thank Boris Dubrovin and Stefano Fantoni for their
invitation and hospitality at the Laboratorio Interdisciplinare per
le Scienze Naturali e Umanistiche at SISSA where this lecture was
first presented. I thank Werner Nahm for providing me with a
copy of his stimulating paper \cite{Nahm} prior to publication. My thanks
also go to S. Yazadjiev who produced a usable file out of my
handwritten notes and to L. Hadjiivanov for a careful reading of the manuscript.

The author acknowledges  partial support from the Bulgarian
National Council for Scientific Research under contract F-828.

\end{document}